\documentclass{osa-article}

\journal{osajournal}
\usepackage{graphicx}
\usepackage{dcolumn}
\usepackage{bm}
\usepackage{comment}
\usepackage{textcomp}
\usepackage{ulem}
\usepackage[]{graphicx,xcolor}
\usepackage{svg}
\usepackage{amsmath}
\usepackage{bbold}
\usepackage[utf8]{inputenc}
\usepackage[T1]{fontenc}
\usepackage[english]{babel}
\usepackage{verbatim}
\usepackage{textcomp}
\usepackage{ulem}
\usepackage[]{graphicx,xcolor}
\usepackage{soul}

\articletype{Research Article}

\begin{document}

\title{Dual-laser self-injection locking to an integrated microresonator}

\author{Dmitry~A.~Chermoshentsev,\authormark{1,2,3,*},
Artem~E.~Shitikov,\authormark{1},
Evgeny~A.~Lonshakov,\authormark{1},
Georgy~V.~Grechko,\authormark{1,2},
Ekaterina~A.~Sazhina,\authormark{2},
Nikita~M.~Kondratiev,\authormark{1},
Anatoly~V.~Masalov,\authormark{1,4}
Igor~A.~Bilenko,\authormark{1,5},
Alexander~I.~Lvovsky\authormark{1,4,6} and Alexander~E.~Ulanov\authormark{1,2, $\dagger$}}

\address{\authormark{1}Russian Quantum Center, Skolkovo, Moscow 143025, Russia\\
\authormark{2}Moscow Institute of Physics and Technology, Dolgoprudny, Moscow Region 141700, Russia\\

\authormark{3}Skolkovo Institute of Science and Technology, Moscow 143025, Russia\\
\authormark{4}Lebedev Physical Institute, Russian Academy of Sciences, 119991, Moscow, Russia\\
\authormark{5}Faculty of Physics, Lomonosov Moscow State University, 119991 Moscow, Russia\\
\authormark{6}Clarendon Laboratory, University of Oxford, Parks Road, Oxford OX1 3PU, UK\\}
\email{\authormark{*}dmitriy.chermoshentsev@phystech.edu}
\email{\authormark{$\dagger$}a.ulanov@rqc.ru}



\begin{abstract}
Diode laser self-injection locking (SIL) to a whispering gallery mode of a high quality factor resonator is a widely used method for laser linewidth narrowing and high-frequency noise suppression. SIL has already been used for the demonstration of ultra-low-noise photonic microwave oscillators and soliton microcomb generation and has a wide range of possible applications. Up to date, SIL was demonstrated only with a single laser. However, multi-frequency and narrow-linewidth laser sources are in high demand for modern telecommunication systems, quantum technologies, and microwave photonics. Here we experimentally demonstrate the dual-laser SIL of two multifrequency laser diodes to different modes of an integrated Si$_3$N$_4$ microresonator. Simultaneous spectrum collapse of both lasers, as well as linewidth narrowing and high-frequency noise suppression , as well as strong nonlinear interaction of the two fields with each other, are observed. Locking both lasers to the same mode results in a simultaneous frequency and phase stabilization and coherent addition of their outputs. Additionally, we provide a comprehensive dual-SIL theory and
investigate the influence of lasers on each other caused by nonlinear effects in the microresonator.
\end{abstract}

\section{Introduction}

Laser is a ubiquitous tool in various modern science and technology applications. Most of these applications, such as atomic clocks \cite{Chou2011, Newman2019}, astronomy \cite{Suh2019}, high-resolution
spectroscopy \cite{Tamm2000}, ultrafast optical ranging \cite{Trocha2018, Suh2018, Riemensberger2020}, and metrology \cite{Nazarova2008}, require stable and narrow-linewidth lasers. Locking a solid-state or fiber laser to a high-finesse cavity is a common way to achieve stability with sub-Hz linewidth \cite{Alnis2008}. However, most of the existing techniques require bulky optical setups and complicated electronics and only operate under controlled laboratory conditions. Developing compact stabilization apparata is therefore of significance for today's optical technology, especially  for semiconductor diode lasers, which feature small form-factors and low energy consumption, and hence can be used in applications where these features are desirable.


An appealing approach is to stabilize a diode laser via self-injection locking (SIL) to an eigenfrequency of a high-Q whispering gallery mode (WGM) crystalline or integrated microring resonator. SIL can be based on the resonant Rayleigh scattering on  volume and surface nonidealities of the  microresonator, which back reflects part of the input radiation to the laser. Backward wave acts as feedback for laser stabilization. 

Since the first demonstration in 1998 \cite{VASSILIEV1998305}, SIL has attracted increasing interest and is being actively studied. The SIL-based sub-Hz lasers and ultra-low-noise photonic microwave oscillators were implemented with crystalline microresonators in 2015 \cite{liang2015ultralow, Liang2015v1}. A comprehensive theory of SIL was developed in Ref.~\cite{Kondratiev2017} and was applied to optimize the experimental parameters in Ref.~\cite{Galiev20}. 

Of particular appeal is SIL in the integrated setting, as it addresses the requirements of form-factor and energy efficiency especially well. The current material of choice for integrated microresonators is silicon nitride ($\mathrm{Si}_3\mathrm{N}_4$) thanks to its low losses, absence of two-photon absorption, fabrication repeatability, and high Kerr nonlinearity \cite{Pfeiffer2016, Pfeiffer2018, Liu2021}. Moreover, Si$_3$N$_4$ permits hybrid integration with InP, offering fully on-chip narrow-linewidth lasers \cite{Huang19,Boller2020}. To date, SIL to integrated Si$_3$N$_4$ resonators have been used for the formation of microcomb solitons, and the development of sub-Hz lasers \cite{Pavlov2018, Raja2019, Kondratiev2020, Lobanov2020, Li2021}. Furthermore, butt-coupling of a DFB diode to an integrated high-Q $\mathrm{Si}_3\mathrm{N}_4$ microresonator was used as a compact SIL-based soliton source \cite{Voloshin2021}. 


Previously, all investigations of SIL considered the interaction of a single laser diode with one or several eigenmodes of a high-Q resonator. 
However, the coupling of two or more laser diodes to different resonances of a single microresonator with nonlinearity
--- a phenomenon to which we refer as dual-SIL ---
 opens up new opportunities for both practical applications and fundamental research. For example, such systems can be used for the generation of frequency combs \cite{Hu2017, Wang2016, Wen2019, Wen2019a} or development of high-power and compact on-chip photonic RF-generators \cite{Liu2020, Khan2010}. Furthermore, dual-SIL 
can be used to build a compact  
optical parametric oscillator \cite{Okawachi2015, Okawachi2016, Vaidya2020, Arrazola2021}, which is a primary tool for various quantum applications such as random number generation \cite{Okawachi2016}, quadrature squeezing \cite{Zhao2020}, Gaussian boson sampling \cite{Arrazola2021} or fully-optical simulation of the Ising model \cite{Okawachi2020}.

Together with its technological appeal, dual-SIL comes with a significant experimental challenge
arising from the complex nonlinear dynamics between the two lasers and the microresonator. Even in the case of single-laser SIL, exciting a microresonator mode results in nonlinear optical effects that shift its frequency. In dual-SIL, the situation is even more complicated because the resonance shift caused by one of the lasers applies to all of the microresonator modes, and hence influences the other laser locking. 


In this paper, we present a theoretical and experimental study of simultaneous SIL of two laser diodes to eigenmodes of a single microring resonator with nonlinearity. First, we develop a model, which  describes this phenomenon and its behavior under varying  experimental parameters. Second, we experimentally implement dual-SIL of two multifrequency laser diodes to a single microresonator. We investigate the cases when the lasers are locked to two different modes and the same mode of the microresonator. The locked lasers exhibit spectral collapse into kHz-wide lines with low phase noise. We show that the frequencies of both lasers can be tuned inside the dual-SIL stability region and that the lasers become nonlinearly coupled through a microresonator. 
This nonlinear coupling results in a four-wave mixing process, producing a comb-like spectral structure if the same family of high-Q microresonator modes is used for the locking. 
Finally, in the case of dual-SIL to a single microresonator mode, we observe coherent adding of the laser outputs, resulting in increasing total output power.


\section{Theoretical model of dual-SIL}

\begin{figure*}
\centering
\includegraphics[width=0.7\linewidth]{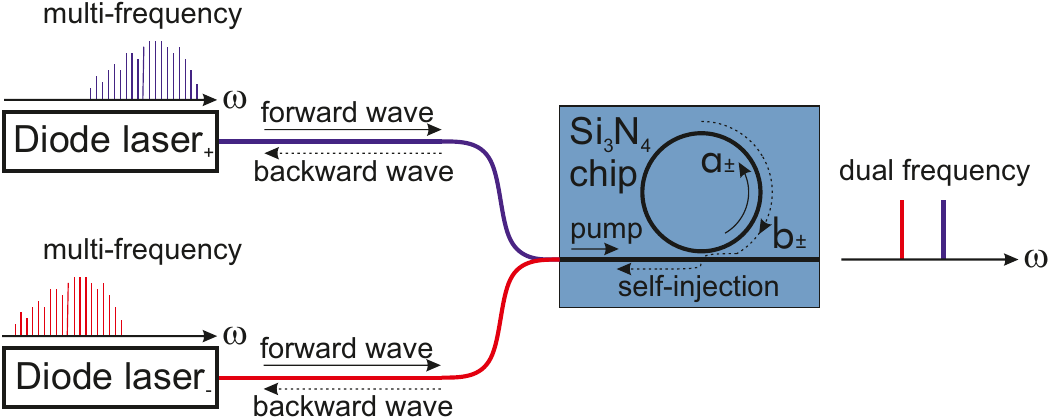}
\caption{Dual-SIL concept. Two laser diodes are coupled to an integrated high-Q microresonator. Both lasers are simultaneously self-injection locked to different frequency modes of the microresonator resulting in a stable and narrow-linewidth bichromatic output.}
\label{fig:concept}
\end{figure*}

We consider the interaction of two independent laser diodes with a single on-chip microresonator (Fig.~\ref{fig:concept}). The excitation of the microresonator mode by the laser produces the backscattered wave, which, in turn, may modify the laser's emission frequency $\omega^{\rm eff}_{\pm}$ with respect to that of the free-running laser $\omega^{\rm{d}}_{\pm}$, where ``$+$" and ``$-$" denote the higher- and lower-frequency lasers, respectively. 
We start the analysis of this dynamics by considering the system of coupled-mode equations for normalized amplitudes of forward $a_{\pm}$ and backward $b_{\pm}$ waves inside the microresonator (Fig.~\ref{fig:concept}): 
\begin{equation}
\begin{aligned}
\label{eq:rate_equation_coupled}
    \dot{a}_{\pm} &=-(1-i\zeta_{\pm})a_{\pm}+i\Gamma_{\pm} b_{\pm}+ia_{\pm}(|a_{\pm}|^2 + 2|a_{\mp}|^2) + 2i\alpha_{x}a_{\pm}(|b_{\pm}|^2 + |b_{\mp}|^2) +f_{\pm}, \\
    \dot{b}_{\pm} &=-(1-i\zeta_{\pm})b_{\pm}+i\Gamma_{\pm} a_{\pm}+ib_{\pm}(|b_{\pm}|^2 + 2|b_{\mp}|^2) + 2i\alpha_{x}b_{\pm}(|a_{\pm}|^2 + |a_{\mp}|^2). &&
\end{aligned}
\end{equation}
These equations are written in the rotating reference frames of the respective pump frequencies. The first term describes the normalized linear losses and frequency detunings $\zeta_\pm=2(\omega^{\rm eff}_\pm-\omega_\pm)/\kappa$ between the lasers' effective emission frequencies $\omega^{\rm eff}_\pm$ and corresponding cold microresonator modes' resonance frequencies $\omega_\pm$
, where $\kappa$ = $\kappa_{c}$ + $\kappa_{i}$ is the total microresonator linewidth, which includes the intrinsic loss $\kappa_{i}$ and external coupling rates $\kappa_{c}$. The second term of Eqs.~\eqref{eq:rate_equation_coupled} determines the influence of the backscattered waves on the forward one and vice versa with the normalized forward-backward wave coupling rates of the microresonator modes $\Gamma_\pm$. The third and fourth terms describe the nonlinear interactions (both self- and cross-phase modulation) with $\alpha_{x}$ being the spatial overlap between the forward and backward modes  which is assumed to be 1 for the modes of the same polarization \cite{Kondratiev2020}. Finally, $f_\pm$ are the normalized input laser diodes' amplitudes, hereafter referred to as the pump. The time and cavity field amplitudes are renormalized with respect to the resonator width and nonlinearity, respectively \cite{Herr2014, Kondratiev2017}.

The stationary solution of Eqs.~\eqref{eq:rate_equation_coupled} connects the amplitudes of the forward $a_{\pm}$ and backward $b_{\pm}$ waves, and consequently, determines the complex reflection coefficients of the microresonator for the given detunings $\zeta_\pm$. Nonlinear effects inside the microresonator lead to the dependence of these coefficients on field intensities inside the ring. Since the backreflection affects the lasers' output frequencies, both lasers become  coupled through these nonlinearities. We emphasize that this coupling is present although the lasers are quite far from each other spectrally, i.e. the back reflection of one laser does not directly affect the other diode and vice versa.

The treatment of this complex nonlinear dynamics simplifies if we introduce modified detunings and backscattering coefficients 
\begin{equation}\label{eq:barred}
     \bar{\zeta}_{\pm} =\zeta_{\pm} + \delta \zeta_{\pm}; \quad 
    \bar{ \Gamma}^{2}_{\pm} = \Gamma^2_{\pm} + \delta \Gamma^{2}_{\pm}
\end{equation}
where the correction terms
\begin{equation}
\label{eqs:shift_and_detuning}
\begin{gathered}
\delta \zeta_{\pm} = \frac{2\alpha_{x} + 1}{2}(|a_{\pm}|^2 + |b_{\pm}|^2) + (\alpha_{x} + 1) (|a_{\mp}|^2 + |b_{\mp}|^2;\\
\delta \Gamma_{\pm} = \frac{2\alpha_{x} - 1}{2}(|a_{\pm}|^2 - |b_{\pm}|^2) + (\alpha_{x} - 1)(|a_{\mp}|^2 - |b_{\mp}|^2).\\ 
\end{gathered}
\end{equation}
represent the nonlinear effects. 
The resulting expression for the microresonator reflection spectrum in terms of the modified detunings $\bar\zeta_\pm$ and modified coupling rates  $\bar\Gamma_\pm$ replicates that in the absence of nonlinearities (see Supplemental Section \textcolor{blue}{1} for detail). 

One can hence apply the linear theory of Kondratiev {\it et al.} \cite{Kondratiev2017} to describe the interaction of each laser with the corresponding mode of the microresonator in the modified coordinates.
This theory studies the interaction of the backreflected wave with a gain medium inside a laser cavity to establish a connection  between the free-running lasers' emission frequencies $\omega_\pm^d$ and the nearest cold microresonator resonance $\omega_{\pm}$. The relation depends on the laser-diode and microresonator parameters and the phase delay related to the distance between the laser and the microresonator. 
The result for the linear case needs to be modified to take into account the nonlinear shift $\delta \zeta_{\pm}$ of the microresonator frequencies;  specifically, the normalized free-running  laser detunings $\xi_{\pm} = 2(\omega^{\rm d}_\pm - \omega_{\pm})/\kappa$ must be adjusted by the same amount: $\bar{\xi}_{\pm} = \xi_{\pm}  + \delta \zeta_{\pm}$. The relations between $\bar{\xi}_{\pm}$ and $\bar{\zeta}_{\pm}$ then take the following form \cite{Voloshin2021}:
\begin{equation}\label{xi_eq}
\bar{\xi}_{\pm} = \bar{\zeta}_{\pm} + \frac{K_{\pm}}{2}\frac{2 \bar{\zeta}_{\pm}\cos{\bar{\psi}_{\pm}} + (1 + \bar{\Gamma}^2_{\pm} - \bar{\zeta}^2_{\pm})\sin{\bar{\psi}_{\pm}}}{(1+\bar{\Gamma}^2_{\pm}-\bar{\zeta}_{\pm}^2)^2+4\bar{\zeta}^{2}_{\pm}},
\end{equation}
where $K_{\pm} = 8\kappa_{c}\Gamma_{\pm}\kappa^{\rm do}_{\pm}/\kappa^2$ are zero-detuning stabilization coefficients which characterize the backscattering feedback with $\kappa^{\rm do}_{\pm}$ being the effective output beam coupling rates of the lasers. The phase delays are given by $\bar{\psi}_{\pm} = \bar{\psi}^0_{\pm}+\kappa\tau_\pm\zeta_\pm/2$, where $\tau_\pm$ are the round-trip times between the corresponding laser and the microresonator, while $\bar{\psi}^0_{\pm}$ are constant alignment-dependent phases, which are assumed zero in this treatment (optimal locking case as per Ref.~\cite{Galiev20}).
 
\begin{figure*}
\centering
\includegraphics[width=1\linewidth]{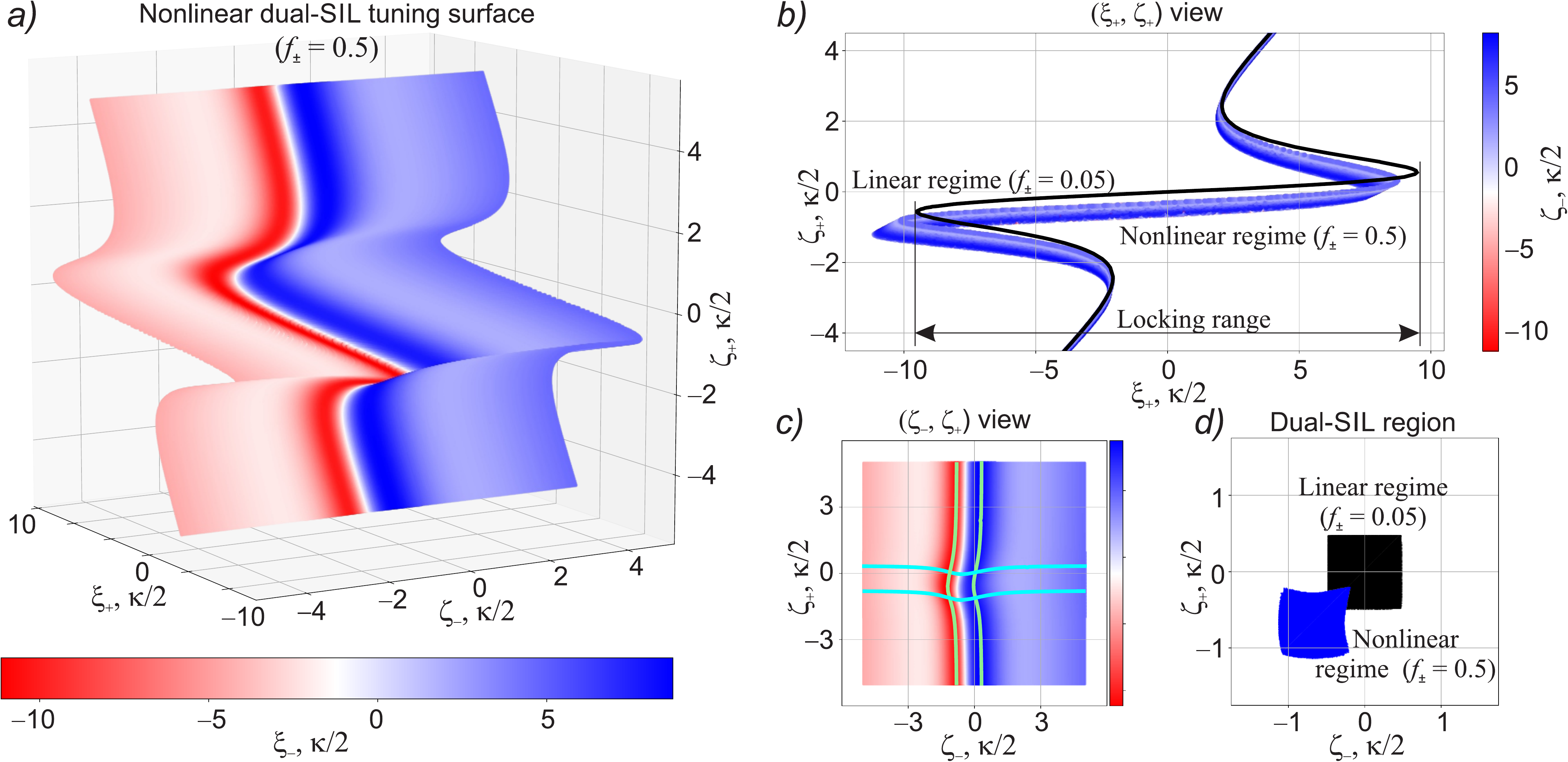}
\caption{Tuning surface of the dual-laser SIL in different views (a-c) and a locking region diagram (d). The surface is calculated for normalized pump amplitudes $f_+=f_-=0.5$, at which moderate nonlinear effects are present. All model parameters are gathered in  Supplementary Table \textcolor{blue}{1}. Color in (a--c) represents the ``$-$" laser intracavity detuning $\xi_{-}$. The black curve in panel (b) and black square in (d) correspond $f_\pm=0.05$, at which nonlinear effects are insignificant. The aqua and light-green curves correspond to the ``$+$" and ``$-$" laser individual locking regions.}
\label{fig:stationary_nonlinear}
\end{figure*}

Equations \eqref{eq:rate_equation_coupled}--\eqref{xi_eq} can be solved numerically (see Supplemental Section \textcolor{blue}{1} for detail) and plotted in the coordinates $(\xi_{+}, \xi_{-}, \zeta_{+}, \zeta_{-})$. As a result, a 4D dual-SIL tuning surface is obtained [Fig. \ref{fig:stationary_nonlinear}(a)]. 

To simplify the interpretation of this surface, we show it in the  $(\xi_+, \zeta_+)$ view in Fig.~\ref{fig:stationary_nonlinear}(b). The black curve in this figure is obtained for $f_+ = f_- = 0.05$ and approaches the weak-pump limit governed by the single-laser linear SIL theory, described by Eq.~\eqref{xi_eq} with $\bar\zeta_\pm\to\zeta_\pm$ and $\bar\Gamma_\pm\to\Gamma_\pm$. In this limit, the two lasers have no effect on each other. At large detuning from microresonator resonances, the lasers are in the free-run regime, so $\zeta_+ = \xi_+$ or $\zeta_- = \xi_-$. Close to resonance, SIL ensues, characterized by the plateau $\partial\zeta_{\pm}/\partial\xi_{\pm}\ll 1$ \cite{Kondratiev2017}.

The colored curves in Fig.~\ref{fig:stationary_nonlinear}(b) corresponds to pump amplitudes $f_+ = f_- = 0.5$, at which nonlinear effects are significant. 
The locking regions (tuning curves' plateaus $\partial\zeta_+/\partial\xi_+\ll 1$) are shifted to the lower frequencies, which is a result of the redshift of microresonator resonances caused by the nonlinearity. The overall shift is due to the nonlinearity associated with the "$+$" laser. The nonlinear effect of the "$-$" laser is manifested by additional shifts of the surface for different $\zeta_-$ values, the latter are shown by different hues. When this laser is closer to the resonance (lighter hues), the power inside the microresonator is higher, so the shift is more significant. The cross-influence of the lasers is also evident in the ($\zeta_-, \zeta_+$) view [Fig.~\ref{fig:stationary_nonlinear}(c)], manifesting as a red shift of the high color gradient region (corresponding to the locking regime of the second laser $\partial\zeta_-/\partial\xi_-\ll1$) in the central region of the plot.

The purple and aqua colors in Fig.~\ref{fig:stationary_nonlinear}(c) show the locking regions of the two lasers, whose boundaries are defined by $\partial\zeta_\pm/\partial\xi_\pm = \infty$. Their intersection determines the dual-SIL region,  displayed by the black and blue areas in Fig.~\ref{fig:stationary_nonlinear}(d) for the weak and strong pumps, respectively. The square shape of the black area is a manifestation of the lasers' mutual independence in the linear case. The blue area's overall red shift and the red-sided curvature of its boundaries represent the self-phase and cross-phase modulation of the fields inside the microresonator by the two lasers.

Generally, dual-SIL requires that not only the derivatives of each field's effective frequency with respect to the corresponding free-running frequency be small ($\partial\zeta_{\pm}/\partial\xi_{\pm}\ll1$), but also that the same condition is satisfied for the cross-derivatives $\partial\zeta_{\pm}/\partial\xi_{\mp}\ll 1$. In the case of moderate pump intensities considered here, the latter condition has little effect provided that the former one is satisfied. However, for even stronger pumps $f_\pm\gtrsim 1$, both conditions must be tested separately when determining the dual-SIL region.

\section{Experimental results}

We study a Si$_3$N$_4$ integrated microring resonator with the free spectral range FSR = 999 GHz, supplied by LIGENTEC. We select the vertically polarized consecutive microresonator eigenmodes located at 1532, 1540 and 1548 nm. These modes are chosen because they are inside the frequency range of the laser diodes  used in the experiment. The full microresonator linewidths at these wavelengths are 1250, 1500 and 1550 MHz, respectively. 
More details on the microresonator modes parameters may be found in Supplementary Table \textcolor{blue}{S2}. 

\subsection{Experimental setup}

\begin{figure*}[htbp]
\centering
\includegraphics[width=\linewidth]{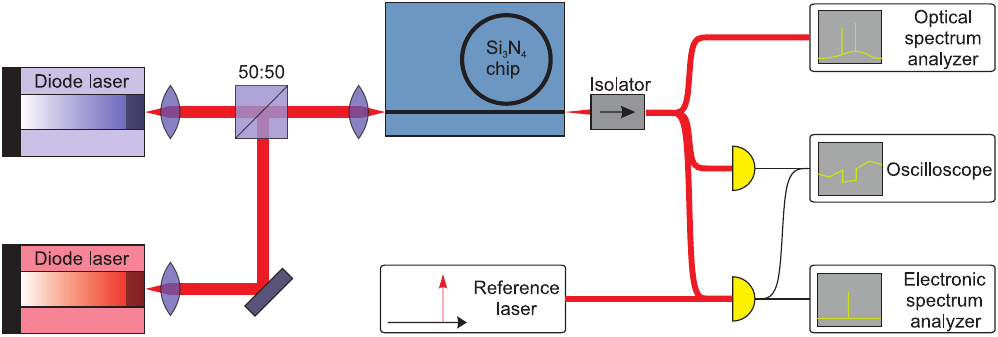}
\caption{Experimental setup. Eigenmodes of the on-chip microring resonator are excited by two multifrequency Fabry-Perot laser diodes. Transmitted light is collected by a lensed fiber. }

\label{fig:setup}
\end{figure*}

For the experimental observation of dual-SIL, we use a setup shown in Fig.~\ref{fig:setup}. Two multi-frequency Fabry-Perot laser diodes (Seminex) mounted on three-axis stages are used as pumps. The first laser diode (referred to as ``blue")  is centered around 1530 nm has an intrinsic intermode distance of $0.3$ nm and output power $\approx 19$ mW; the second (``red")  diode with an intermode distance of $0.15$ nm and output power $\approx 18$ mW is centered around 1545 nm. 

Collimated beams from both lasers are combined on a symmetric beamsplitter and coupled to the Si$_3$N$_4$ chip with an estimated efficiency of 40\%. This corresponds to the pump amplitudes $f_{\rm blue}=f_+ = 0.442$ and $f_{\rm red}=f_-= 0.388$, meaning that we work in the regime at which moderate nonlinear effects are observed [Fig.~\ref{fig:stationary_nonlinear}(d)].

The temperature of the chip is stabilized by a thermo-electric controller. 
The transmitted light at the end of the waveguide is collected with a lensed single-mode fiber and subjected to three measurements: (1) optical spectrum analysis, (2) acquisition of the total transmission with a photodetector and (3) heterodyne detection with the help of a reference laser (Toptica CTL 1550) and a fast photodetector. 

The laser diodes do not have optical isolators in order to enable the Rayleigh backscattering to provide the optical feedback. The rough tuning of the lasers is achieved by changing their temperatures and the fine tuning by modulating their currents. When one of the laser lines is swept through a microresonator resonance, we observe collapse of the emission spectra into a single line \cite{galiev2018spectrum} accompanied by a dip in transmission due to the coupling to the microresonator. These two observations signify single-laser SIL. Importantly, similar signatures have been observed when a laser locked itself to the weak Fabry-Perot resonator formed by the chip front and back facets. The two resonators could be distinguished by the transmission dip magnitude, which was much more significant for the microresonator resonances, and by the wavelength of the resultant emission being consistent with the microresonator modes.

We analyze two different cases: first, when the lasers excite the microresonator modes separated by 2 microresonator FSR intervals (at 1532~nm and 1548~nm respectively), second, when both lasers excite the same mode at 1540~nm.

\subsection{Dual-SIL of lasers separated by 2 FSRs}
We adjust the temperature and current of the "blue" and "red" diodes to center their emission around 1532 and 1548 nm respectively. The experimental challenge in achieving dual-SIL, as discussed above, arises due to the nonlinear effects: when one of the lasers excites a microresonator mode, the spectrum of the latter shifts, thereby possibly knocking the laser out of lock. This was addressed by iterative tuning of lasers to their respective microresonator modes.  

\begin{figure*}[htbp]
\centering
\includegraphics[width=\linewidth]{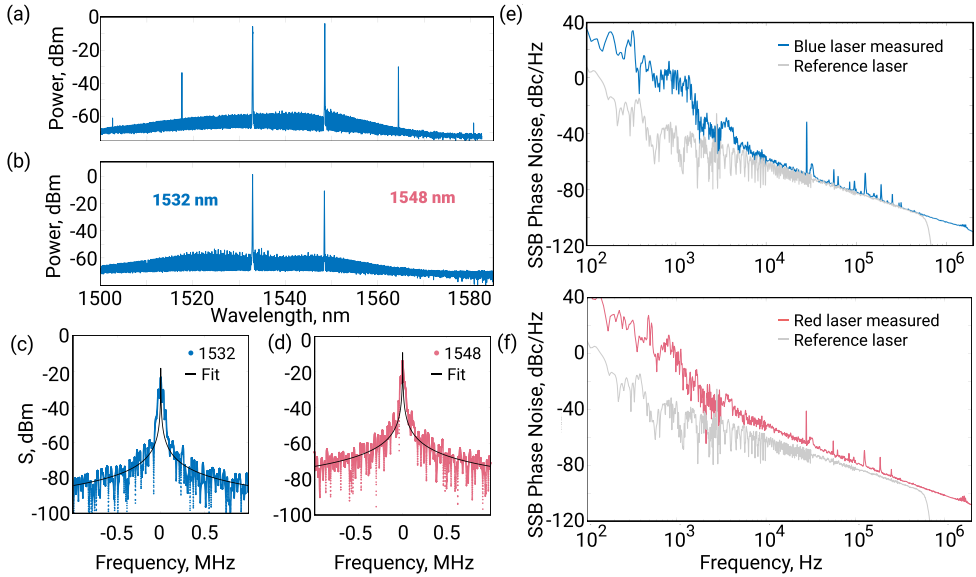}
\caption{Spectral properties of the dual-laser SIL to frequency modes separated by a 2-FSR interval. a) Optical spectrum in the regime when a frequency comb is generated in a four-wave mixing process. b)  Optical spectrum with frequency comb generation suppressed. c--d) Heterodyne signal spectra of the 1532 nm (c) and 1548 nm (d) lasers.
e--f) Single-sideband phase noise spectral density of the 1532 nm (e) and 1548 nm (f) lasers. The grey line is the phase noise level of the reference laser in the absence of signal.
} 
\label{fig:spectral_properties}
\end{figure*}

The optical spectrum of the two lasers under dual-SIL is displayed in Fig.~\ref{fig:spectral_properties}(a,b). A frequency comb
was generated due to four-wave mixing [Fig. \ref{fig:spectral_properties}(a)] with the lines separated by 2 FSR. In the case of a bichromatic pump, this process is thresholdless,  but requires certain detunings of the pump lasers in order for the comb lines, corresponding to linear combinations of the laser frequencies, to match microresonator eigenfrequencies. 
However, in order to characterize the dual-SIL, we chose such powers and detunings at which additional comb lines do not emerge [Fig. \ref{fig:spectral_properties}(b)]. 

The spectrum of the heterodyne detection signal is presented in Fig.~\ref{fig:spectral_properties}(c,d). Lorentz approximation of the spectrum gives the instantaneous laser linewidths: 3 kHz for the laser at 1532 nm and 4.5 kHz for the one at 1548 nm.
Further, we acquire and process the in-phase and quadrature components (I/Q-data) of the heterodyne signals to extract the single-sideband (SSB) spectral density of the phase noise [Fig.~\ref{fig:spectral_properties}(e,f)]. The gray line is a phase noise level of the reference laser. We see that the phase noise of our lasers is significantly higher than that of the reference laser only at low frequencies. The reason for the excess noise is technical (acoustics and temperature). A particularly significant phase noise source is the fluctuations in the optical path length between the lasers and the chip. This can be addressed e.g.~by means of butt coupling of the laser diode and isolating the setup.

To investigate the lasers' dynamics within the dual-SIL zone, we lock both laser diodes to the corresponding resonator modes. Then we apply a 40~Hz triangular current modulation to one of the laser diodes and perform heterodyne measurements of both signals. The modulation amplitude is set to 50~mA, which approximately corresponds to the 11 and 5.5~GHz of free-run frequency shift for the blue and red laser respectively. This is significantly larger than the $1.35\pm0.15$ GHz resonator mode linewidth, so the modulation drives the lasers across an entire microresonator line.  
\begin{figure*}[htbp]
\centering
\includegraphics[width=\linewidth]{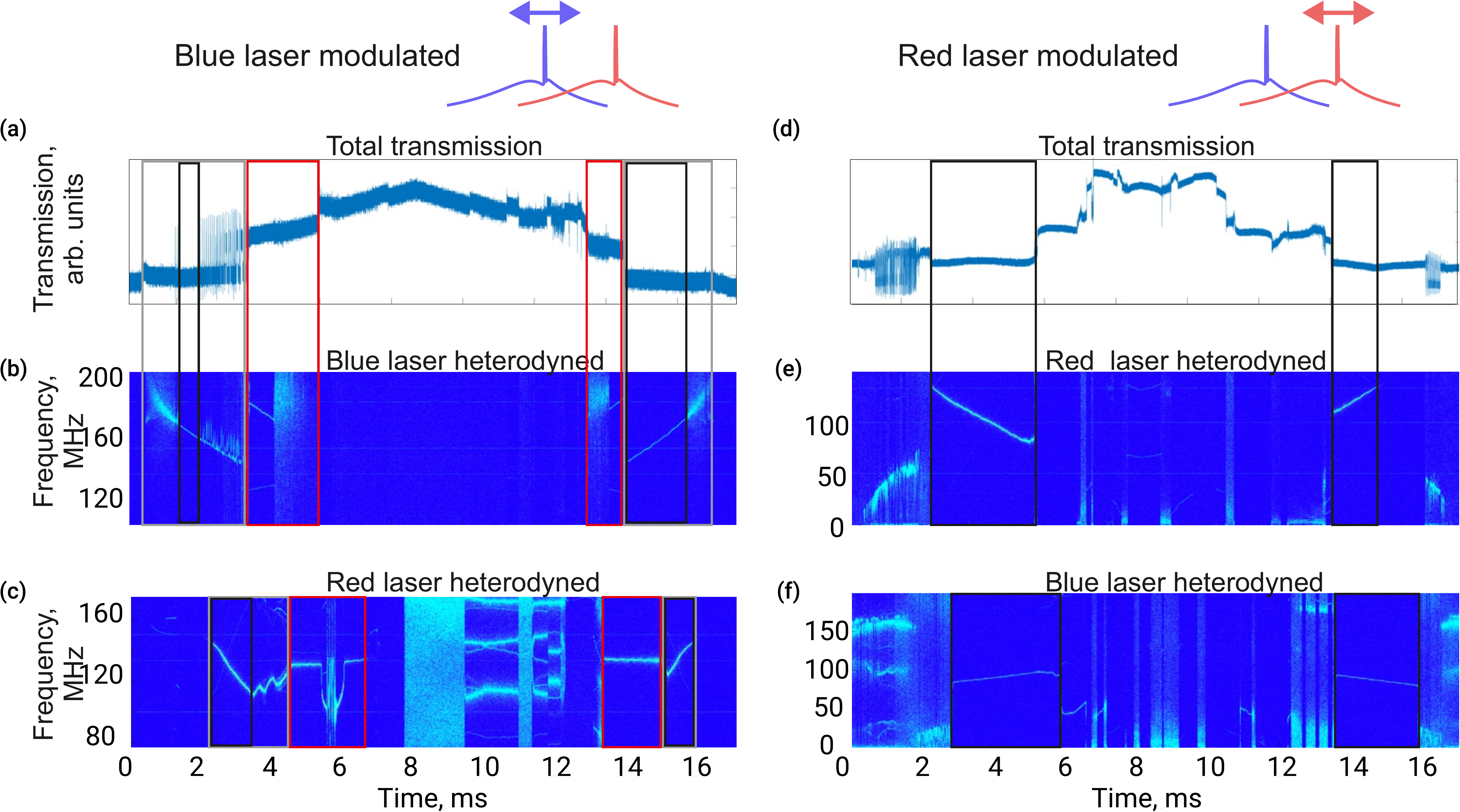}
\caption{Dynamics of the dual-SIL effect during a single forward and backward laser frequency sweep. The blue laser is swept in the left column and red laser in the right column. The transmission traces (a), (d) are shown together with the instantaneous spectrograms (b,c,e,f). The acquisition of the spectrograms in (b) and (e) was syncronized with (a) and (d), respectively, while those in (c) and (f) were acquired in separate runs. The black rectangles correspond to the stable and unstable dual-SIL regimes, respectively. The red rectangles relate to the regime when only the red laser is locked to the correct eigenmode.}
\label{fig:spectrogram}
\end{figure*}

Figure \ref{fig:spectrogram} (a,d) shows the total transmission signal. Dips in the transmission correspond to the excitation of various on-chip modes, including the fundamental and higher-order microresonator modes and the Fabry-Perot cavity formed by the chip facets. The deepest ones (gray shading in Fig.~\ref{fig:spectrogram}a,d) correspond to the simultaneous excitation of both fundamental microresonator modes under study. Black rectangles highlight stable SIL while grey rectangles show transition regions, in which nonlinear and thermal effects drive the blue laser in and out of lock. The areas inside red rectangles [Fig.~\ref{fig:spectrogram}(a)] correspond to the case when the red laser excites the resonator fundamental frequency mode (1548 nm), while the blue laser does not. 

Figure \ref{fig:spectrogram} (b,c,e,f) shows time-dependent spectrograms of the heterodyne signal for four cases: the blue laser being modulated while the LO is at the blue (b) or red (c) laser frequency and the red laser is swept while the red (e) or blue (f) lasers are heterodyned. Single narrow lines are observed inside the dual-SIL region. The laser signal frequency change in these regions is about a few tens of MHz, which is two orders of magnitude less than the free-running laser frequency change corresponding to the same laser current variation. The transmission traces [Fig.~\ref{fig:spectrogram} (a,d)] and the spectrograms in Fig.~\ref{fig:spectrogram} (b,e) have been acquired sinchronously. The spectrograms in Fig.~\ref{fig:spectrogram} (c,f) were acquired in a separate measurement, which required re-tuning of the reference laser. The phase drifts that take place during this re-tuning are responsible for a slight difference in the features observed in these two spectrograms, however the general behaviour remains the same.   

In addition to dual-SIL to microresonator fundamental modes, we also regularly observed situations in which one or both of the lasers were locked to a microresonator mode of a different family or a Fabry-Perot modes of the chip. An example of locking to different family modes is presented in the Supplementary Section \textcolor{blue}{2}. This locking regime was sometimes even more stable than dual-SIL because the nonlinear interaction between the two lasers is reduced or absent due to orthogonality of spatial transverse profiles of modes from different families. However, as discussed previously, this absence of interaction precludes many important applications of dual-SIL.

\subsection{Coherent addition}
\begin{figure*}[htbp]
\centering
\includegraphics[width=\linewidth]{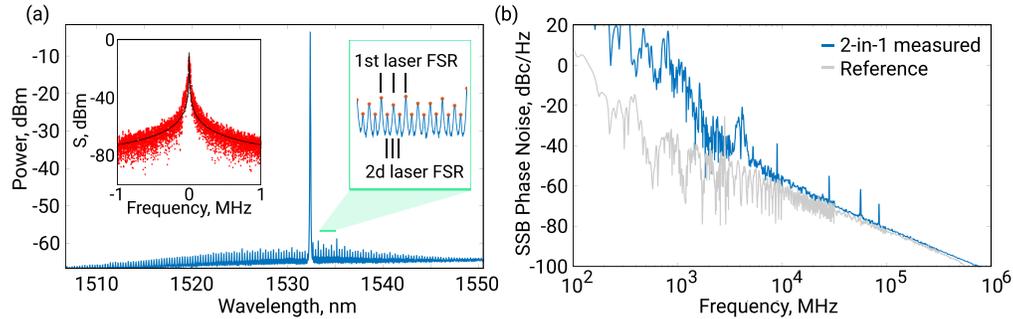}
\caption{Spectral properties of the dual-laser SIL to the same resonance mode. (a) Spectrum acquired with the optical spectum analyzer. The left inset shows the heterodyne signal spectrum approximated with the Lorentz profile. The right inset presents the supressed longitudinal modes of the lasers separated by the free-spectral ranges of their cavities. (b) Single sideband phase noise spectral density.
}
\label{fig:2in1}
\end{figure*}

We also investigate a case when both lasers are simultaneously locked to a single microresonator eigenmode at 1540 nm.
We observe a single line in the optical spectrum [Fig.~\ref{fig:2in1}(a)] and we also checked that there were no beat signals in the full 25 GHz bandwidth of the fast photodetector.  This suggests that both lasers were frequency locked to the same mode. We can characterize this regime as mutual injection locking of the two lasers to each other \cite{liu2019optical}. We present further discussion of this matter in Supplementary Section \textcolor{blue}{3}.

As in the previous section, the heterodyne signal spectrum and I/Q data  were measured. The line spectrum approximated with the Lorentz profile gives an instantaneous linewidth estimation of 4 kHz [Fig.~\ref{fig:2in1}(a), left inset]. The I/Q data [Fig.~\ref{fig:2in1}(b)] exhibits the phase noise at a similar level to the case of dual-SIL to resonator modes separated by 2 FSR, which means that the phases of both lasers synchronized. In addition, we found an increase in the total output power of the lasers. To show this, we tapped a small fraction of the laser radiation before the chip and monitored it with the power meter. The powers measured were 0.43~mW and 0.51~mW for each of the lasers separately in the single-SIL case while the combined power in the dual-SIL regime was 1.5~mW. We attribute this effect to a stronger resonant feedback wave entering each laser.

\section{Outlook}
Possible directions for future research include multi-frequency and narrow-linewidth laser sources, integrated low-threshold microcomb generators and high output power photonic microwave generators, which have multiple applications in telecommunications, spectroscopy, and metrology. Moreover, this setup is ideally suited for quantum light applications since it can be used as a fully integrated degenerate optical parametric oscillator.
\section*{Note added} 
While our work was being prepared for publication, we became aware of an experiment on self-injection locking of two diode lasers to two different modes of an integrated ring microresonator, achieving a linewidth of 17 kHz \cite{Jiang:21}. This work was done in a different setting: the locking was implemented using a forward-propagating wave outcoupled from the microresonator via an additional waveguide, which was coupled back into the laser using a circulator. Single-mode VCSELs, rather than high-power multimode diode lasers, have been used. Importantly, Ref.~\cite{Jiang:21} shows no nonlinear effects between the two fields, which, as discussed above, are essential for many practical applications of dual-SIL.

\begin{backmatter}
\bmsection{Acknowledgements}
We thank M. Karpov for fruitful discussions, N. Dmitriev for assistance with the experiment, and I. Stepanov for performing COMSOL simulations. The theoretical part of the reported study was funded by RFBR, project number 20-32-90172, experimental research was supported by the Russian Science Foundation, grant 20-12-00344. This work was supported by the Russian Roadmap for Quantum Computing.
\bmsection{Disclosures}
The authors declare no conflict of interests.
\bmsection{Supplemental document}
See Supplement 1 for supporting content. 
\end{backmatter}
\bibliography{bibliography}
\end{document}


\maketitle

\section{Theory of dual-SIL}
To analyse the dual-self injection locking regions we the find stationary solution of Eqs.~(1) of the main text, which we reproduce here:

\begin{equation}
\label{eq:rate_equation_coupled_stationary}
\begin{gathered}
    -(1-i\zeta_{\pm})a_{\pm}+i\Gamma_{\pm} b_{\pm}+ia_{\pm}(|a_{\pm}|^2 + 2|a_{\mp}|^2) + 2i\alpha_{x}a_{\pm}(|b_{\pm}|^2 + |b_{\mp}|^2) +f_{\pm} = 0; \\
    -(1-i\zeta_{\pm})b_{\pm}+i\Gamma_{\pm} a_{\pm}+ib_{\pm}(|b_{\pm}|^2 + 2|b_{\mp}|^2) + 2i\alpha_{x}b_{\pm}(|a_{\pm}|^2 + |a_{\mp}|^2) = 0.
\end{gathered}
\end{equation}
We are interested in the complex reflection coefficient of the microring resonator, which is described in the normalized units as follows  \cite{Gorodetsky:00}:
\begin{equation}
    \label{eq:reflectance_coefficient}
    R_\pm(\omega) = - 2 \frac{\kappa_c}{\kappa}\frac{b_\pm}{ f_\pm}.
\end{equation}
From the second equation of \eqref{eq:rate_equation_coupled_stationary}, the forward amplitude can be expressed as follows:
\begin{equation}
    \label{eq:forward_wave}
    a_\pm = \frac{(1-i\zeta_\pm)-i(|b_\pm|^2+2|b_\mp|^2+2\alpha_x (|a_\pm|^2+|a_\mp|^2))}{i\Gamma_\pm} b\pm.
\end{equation}
We substitute this result into the first equation of Eqs.~(\ref{eq:rate_equation_coupled_stationary}), obtaining
\begin{equation}\label{eq:ref_derivation}
\begin{gathered}
    -i\Gamma_\pm\frac{f_\pm}{b_\pm} = \left((1-i\zeta_\pm)-i(|b_\pm|^2 +  2 |b_\mp|^2+2\alpha_x(|a_\pm|^2+|a_\mp|^2)) \right)\\
    \left(-(1-i\zeta_\pm)+i(|a_\pm|^2+2|a_\mp|^2+2\alpha_x(|b_\pm|^2+|b_\mp|^2))\right) - \Gamma_\pm^2.
\end{gathered}
\end{equation}
In the absence of nonlinearities, this result becomes $-i\Gamma_\pm {f_\pm}/{b_\pm} = -(1-i\zeta_\pm)^2 - \Gamma_\pm^2$. Our goal is to introduce  new variables $\bar{\zeta}_{\pm} = \zeta_{\pm} + \delta \zeta_{\pm}$ and $\bar{ \Gamma}^{2}_{\pm} = \Gamma^{2}_{\pm} + \delta \Gamma^{2}_{\pm}$  such that Eq.~(\ref{eq:ref_derivation}) takes a similar form when written in these variables, i.e. 
\begin{equation}\label{eq:ref_after_subst}
    -i\Gamma_\pm \frac{f_\pm}{b_\pm} = -(1-i\bar{\zeta}_\pm)^2 - \bar{\Gamma}_\pm^2.
\end{equation}
To this end, we equate the right-hand sides of Eqs.~(\ref{eq:ref_derivation}) and (\ref{eq:ref_after_subst}):
\begin{equation}\label{eq:ref_derivation_extended}
    \begin{gathered}
-(1-i\zeta_\pm)^2 + i(1-i\zeta_\pm)\left[|a_\pm|^2+|b_\pm|^2 +2 |a_\mp|^2 + 2|b_\mp|^2 + 2\alpha_x(|a_\pm|^2+|a_\mp|^2+|b_\pm|^2+|b_\mp|^2)\right] +\\ \left[|b_\pm|^2+2|b_\mp|^2+2\alpha_x(|a_\pm|^2+|a_\mp|^2)\right]\left[|a_\pm|^2+2|a_\mp|^2+2\alpha_x(|b_\pm|^2+|b_\mp|^2)\right] - \Gamma_\pm^2 =\\-(1 - i \zeta_\pm)^2 + 2i(1-i\zeta_\pm)\delta\zeta_\pm + \delta\zeta_\pm^2 - \Gamma_\pm^2 - \delta\Gamma_\pm^2 .
    \end{gathered}
\end{equation}
The coefficients of similar powers of $(1-i\zeta_\pm)$ must be equal in both sides of \eqref{eq:ref_derivation_extended}, which yields: \begin{equation}\label{eq:corrections_1}
    \begin{gathered}
    2\delta\zeta_\pm = |a_\pm|^2+|b_\pm|^2  +2|a_\mp|^2+2|b_\mp|^2+2\alpha_x(|a_\pm|^2+|a_\mp|^2+|b_\pm|^2+|b_\mp|^2);\\
    \delta\zeta_\pm^2 - \delta\Gamma_\pm^2 = \left(|b_\pm|^2 +2|b_\mp|^2+ 2\alpha_x(|a_\pm|^2+|a_\mp|^2)\right)\left(|a_\pm|^2+2|a_\mp|^2+2\alpha_x(|b_\pm|^2+|b_\mp|^2)\right).
    \end{gathered}
\end{equation}
From these, the nonlinear corrections are as follows:
\begin{equation}
\label{eqs:shift_and_detuning}
\begin{gathered}
\delta \zeta_{\pm} = \frac{2\alpha_{x} + 1}{2}(|a_{\pm}|^2 + |b_{\pm}|^2) + (\alpha_{x}+1)(|a_{\mp}|^2 + |b_{\mp}|^2); \\ 
\delta \Gamma_{\pm} = \frac{2\alpha_{x} - 1}{2}(|a_{\pm}|^2 - |b_{\pm}|^2) + (\alpha_{x}-1)(|a_{\mp}|^2 - |b_{\mp}|^2) .
\end{gathered}
\end{equation}

Finally, from the \eqref{eq:ref_after_subst} and \eqref{eq:forward_wave} the amplitudes of forward $a_\pm$ and backward $b_\pm$ waves are described as:
\begin{equation}
    \begin{gathered}
    a_\pm = \frac{(1 - i\bar{\zeta} - i\delta\Gamma_\pm)f_\pm}{(1 - i\bar{\zeta}_\pm)^2+\bar{\Gamma}_\pm^2};\\
    b_\pm = \frac{i\Gamma_\pm f_\pm}{(1 - i\bar{\zeta}_\pm)^2 + \bar{\Gamma}_\pm^2}.
    \end{gathered}
\end{equation}
The obtained amplitudes can be substituted into the system (\ref{eq:rate_equation_coupled_stationary}) and the new equations for the corrections $\delta\zeta$ and $\delta\Gamma$ will take the following form:
\begin{equation}\label{zeta_eq}
\begin{gathered}
(\alpha_x+1)\delta\zeta_{\pm}-\frac{2\alpha_x+1}{2}\delta\zeta_{\mp}=\frac{4\alpha_x+3}{4}f_{\mp}^2\frac{1+(\bar\zeta_{\mp}+\delta\Gamma_{\mp})^2+\Gamma_{\mp}^2}{(1+\bar\Gamma_{\mp}^2-\bar\zeta_{\mp}^2)^2+4\bar\zeta_{\mp}^2};\\
(\alpha_x-1)\delta\Gamma_{\pm}-\frac{2\alpha_x-1}{2}\delta\Gamma_{\mp}=\frac{3-4\alpha_x}{4}f_{\mp}^2\frac{1+(\bar\zeta_{\mp}+\delta\Gamma_{\mp})^2-\Gamma_{\mp}^2}{(1+\bar\Gamma_{\mp}^2-\bar\zeta_{\mp}^2)^2+4\bar\zeta_{\mp}^2}.
\end{gathered}
\end{equation}
 This system of equation can be solved numerically with respect to $\delta \zeta_{\pm}$ and $\delta \Gamma_{\pm}$ for a given ($\zeta_{+}, \zeta_{-}$). Finally, the complex reflection coefficient (\ref{eq:reflectance_coefficient}) can be calculated as
 \begin{equation}\label{eq:reflectance_coefficient_final}
    R_\pm(\omega) = - 2 \frac{\kappa_c}{\kappa}\frac{i\Gamma}{(1-i\bar{\zeta}_\pm)^2+\bar{\Gamma}^2}.
 \end{equation}
 By using the reflection coefficient (\ref{eq:reflectance_coefficient_final}) and the approach described in Ref.~\cite{Kondratiev2017}, one computes the relation between the free-running and effective laser frequencies, which is given by Eq.~(3) in the main text. The results of the simulation are shown in Fig.~2 of the main text for the set of parameters listed  in Table \ref{table:stationary_params}. Note that the parameters used for the simulation are chosen slightly different from the actual experimental parameters (Table~\ref{table:modes_params_2d_ring}) for clearer presentation of nonlinear effects.

\begin{table}[ht]
\caption{\label{table:stationary_params} Parameters of the simulation in stationary analysis of dual-SIL}
\vspace{-1em}
\begin{center}

\begin{tabular}{|c|c|c|c|c|c|c|c|c|c|}
\hline FSR&$Q$&$\kappa_{do}/2\pi$ &$\kappa_{c}/2\pi $& $\kappa_{i}/2\pi $ & $K_\pm$ & $\alpha_x$ & $\Gamma_\pm$ &$\tau_s$ &$\psi^\pm_{0}$  \\\hline
1 THZ &$4\cdot 10^{5}$ &$7\cdot 10^3$ MHz& $50$ MHz &$435$ MHz  &$25.5$ & $1.0$ & $0.2$& $1.7\cdot 10^{-3}$ ms & $0$\\\hline
\end{tabular}
\end{center}
\end{table}

\begin{table}[ht]
\caption{\label{table:modes_params_2d_ring} Parameters of modes in the experiments with locking to modes from the same family.
}
\vspace{-1em}
\begin{center}
\begin{tabular}{|c|c|c|c|}
\hline
$\lambda$, nm & $\kappa_{i}/2\pi, MHz$ & $\kappa_{c}/2\pi$, MHz & $\Gamma$ \\\hline\hline
1532.86 & 100 & 1137 & 0.64  \\\hline
1540.62 & 106 & 1396 & 0.59 \\\hline
1548.47 & 108 & 1450 & 0.49 \\\hline
\end{tabular}
\end{center}
\end{table}
\section{Self-injection locking to modes from different families}
We performed an experiment on simultaneous locking of two diode lasers to modes from different families. 

While the ``red'' laser interacted with a fundamental mode at 1548.47 nm, the ``blue'' laser was tuned to a non-fundamental one at 1532.40 nm (while the corresponding fundamental mode was at 1532.86 nm). 
The main distinguishing feature for SIL to two modes of different families was the absence of nonlinear sideband generation due to lack of mode matching.

\begin{figure*}[ht]
\centering
\includegraphics[width=1\linewidth]{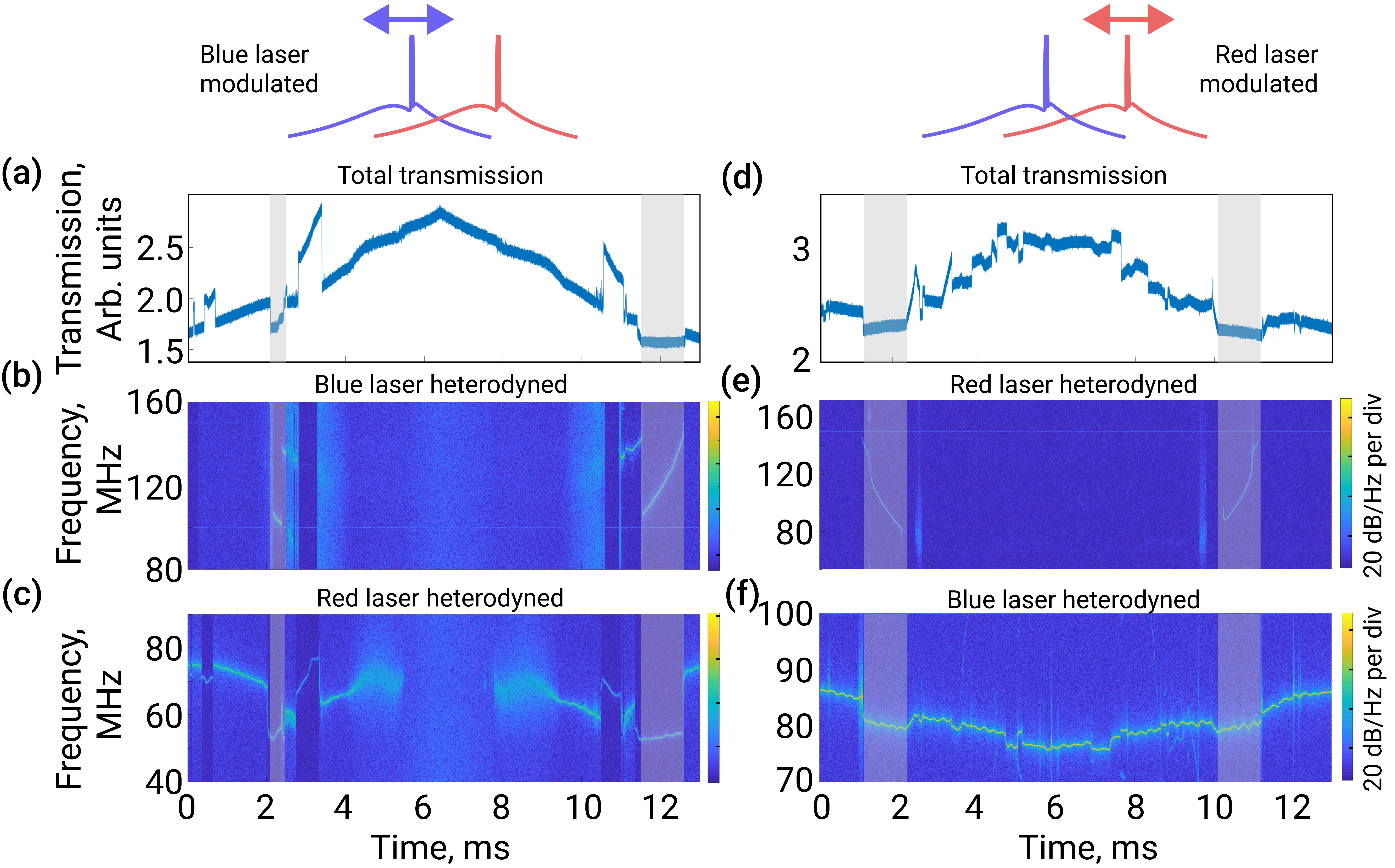}
\caption{Dynamics of the SIL to two modes from different families. The transmission traces (a), (d) and heterodyne spectrograms correspond to frequency sweep of one of the lasers (blue laser in the left column and red laser in the right column). In (b,f), the heterodyne measurement is performed on the same laser as the one being modulated;  in (c,e) on the other laser.}
\label{fig:spectrogramm_differ}
\end{figure*}

For technical reasons, we investigated this regime with a different microring resonator on the same chip, which was however virtually identical to the microresonator used for other experiments.
We measured the transmission traces and spectrograms with the same setup as described in the main text of the manuscript. The results are presented in Fig.~\ref{fig:spectrogramm_differ}.  The major dips in the transmission [grey shadowing in Fig.~\ref{fig:spectrogramm_differ}(a,d)] correspond to the excitation of the resonator modes under study. The shallow dips correspond to the excitation of various Fabry-Perot type modes formed, presumably, by the chip facets.

Similarly to the experiment in the main text, we performed  spectroscopy of both lasers: while one of the lasers was modulated, each of two the fields was subjected to heterodyne measurement in separate experimental runs. When the measurement was performed on the same laser that was modulated, a narrow SIL line was observed only in the resonator mode excitation region [Fig.~\ref{fig:spectrogramm_differ}(b,e)], while otherwise the laser diode had broad multimode spectrum. The signal frequency change in these regions corresponds to the laser frequency shift inside the SIL zone similar to the single-laser SIL \cite{Kondratiev2017}. In contrast, the other laser's narrow SIL was essentially oblivious to the modulation, manifesting virtual absence of interaction [Fig.~\ref{fig:spectrogramm_differ}(c,f)]. The abrupt shifts of the SIL line in the grey shadow regions are likely due to the thermo-optical effect caused by the microresonator mode being excited by the modulated laser in the SIL regime. Slow modulations outside those regions are also likely to the thermal effects related to 
the power change of the modulated laser as the frequency scan is performed by changing the injection current.


\section{Mutual injection locking through the single microresonator mode}
\begin{figure*}[ht]
\centering
\includegraphics[width=0.8\linewidth]{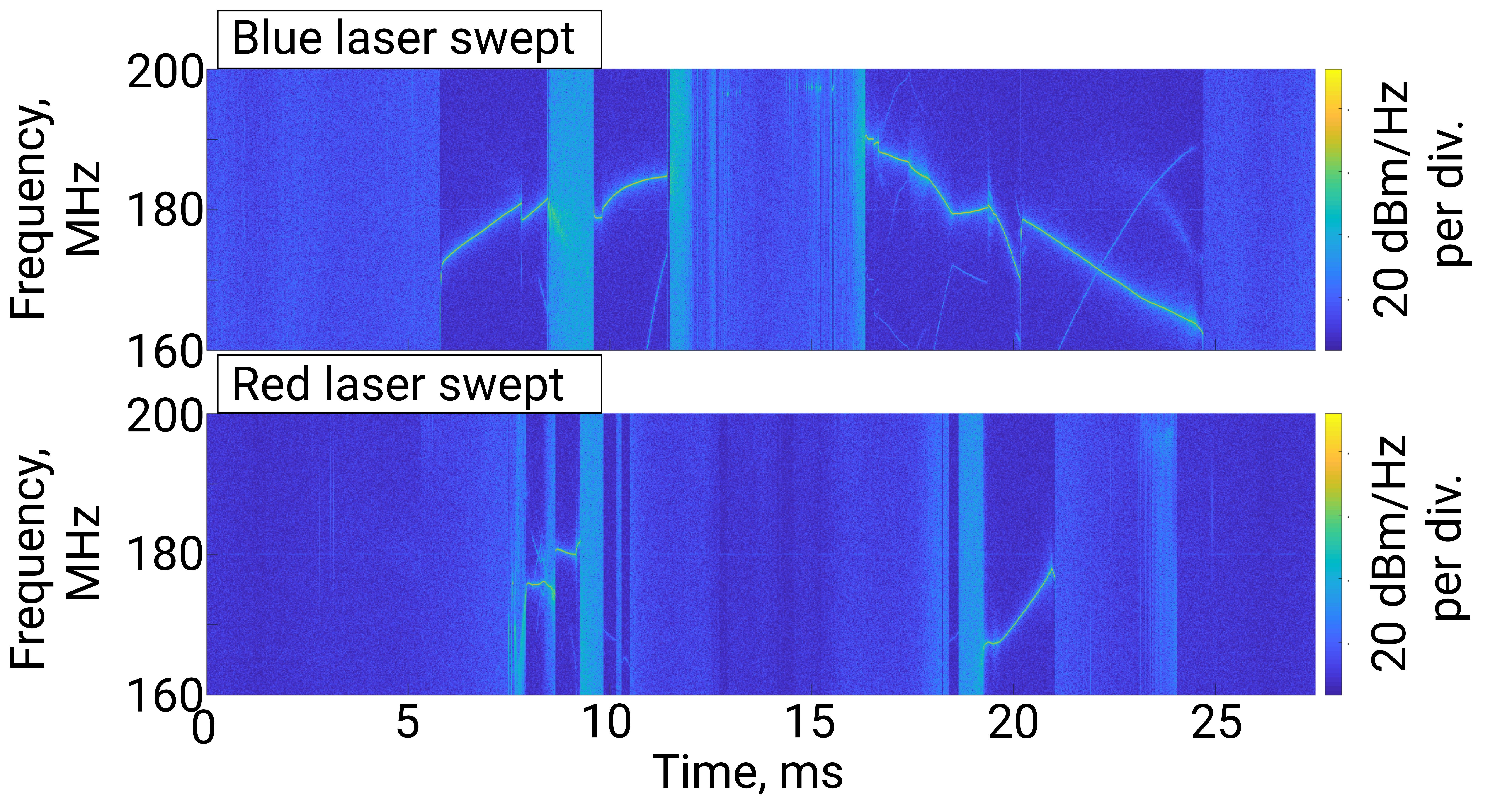}
\caption{Dynamics of the dual-SIL effect in the case of locking to a single mode. The spectrograms correspond to the frequency modulation of each of the lasers, while the local oscillator is at the same frequency.}
\label{fig:spectrogramm_2in1}
\end{figure*}

Fig. \ref{fig:spectrogramm_2in1} shows a detailed study of the case when both lasers are simultaneously locked to a single microresonator eigenmode at 1540 nm. 
Although the lasers operate at the same frequency in this experiment, we still refer to them as ``blue'' and ``red'' following the previous convention. 
An interesting feature of this regime is that the red laser has a stronger influence on the blue one than vice versa. 
Indeed, the locking range of the ``blue" laser is wider then that of the red one, meaning that a more significant perturbation of its free-running frequency is required to knock the system out of lock. At the same time the locking range of the ``red'' laser was similar to its locking range in absence of the ``blue'' one.
The dominant role of the red laser with respect to the blue one was observed consistently in all experimental runs. 
The reasons for this effect are to be investigated. 
We should note that the observed effect is different from the common master-slave injection locking as in our case the coupling of the two lasers to each other was approximately equal. 

\bibliography{bibliography}